\documentclass[conference]{IEEEtran} 

\usepackage{amsmath}
\usepackage{amssymb}
\usepackage{array}
\usepackage{fixmath}
\usepackage{graphicx}
\usepackage{cite}
\usepackage{color}
\usepackage{changes}
\usepackage{algorithm}
\usepackage{algorithmic}
\usepackage{blindtext}
\usepackage{acronym}
\usepackage{soul}
\usepackage{svg}
\usepackage{xprintlen}
\usepackage{nameref}
\usepackage{nicefrac}

\usepackage[font=footnotesize]{caption}
\usepackage[font=footnotesize]{subcaption}

\usepackage{tikz}
\usepackage{filemod}
\usepackage{pgfplots}
\usetikzlibrary{spy,backgrounds}
\usetikzlibrary{arrows,	
	shapes, 
	positioning, 
	fit,	
	backgrounds, 
	mindmap,
	petri,
	intersections, 
	external 
}

\pgfplotsset{compat=1.16}

\DeclareMathAlphabet{\mathbit}{OML}{cmr}{bx}{it}

\newcommand{\B}[1]{{\bf{#1}}}
\DeclareMathOperator{\Transpose}{T}
\DeclareMathOperator{\Hermitian}{H}
\newcommand{\Tr}{{\Transpose}}
\newcommand{\He}{{\Hermitian}}

\DeclareMathOperator{\spann}{span}

\DeclareMathOperator*{\argmax}{argmax}

\newcommand{\eul}{{\text{e}}}
\newcommand{\imj}{{\text{j}}}

\usepackage[numbers,sort&compress]{natbib}

\newcommand\kBob{K_{\rm B}}

\newcommand\kEve{K_{\rm E}}

\newcommand{\GJAL}[1]{{\color{black} #1}}

\acrodef{5G}[5G]{fifth generation}
\acrodef{6G}[6G]{sixth generation}
\acrodef{ADC}[ADC]{analog-to-digital converter}
\acrodef{AO}[AO]{alternating optimization}
\acrodef{AoA}[AoA]{angle of arrival}
\acrodef{AoD}[AoD]{angle of departure}
\acrodef{APN}[APN]{analog precoding network}
\acrodef{ASK}[ASK]{amplitude-shift keying}
\acrodef{AWGN}[AWGN]{additive white Gaussian noise}
\acrodef{BER}[BER]{bit error ratio}
\acrodef{BF}[BF]{beamforming}
\acrodef{BFN}[BFN]{beamforming network}
\acrodef{BS}[BS]{base station}
\acrodef{CB}[CB]{conjugate beamforming}
\acrodef{COMP}[COMP]{covariance OMP}
\acrodef{CSI}[CSI]{channel state information}
\acrodef{DAC}[DAC]{digital to analog converter}
\acrodef{DBS}[DBS]{distance-based scheduling}
\acrodef{DCS}[DCS]{digital communication system}
\acrodef{DCOMP}[DCOMP]{dynamic COMP}
\acrodef{DFT}[DFT]{discrete Fourier transform}
\acrodef{DL}[DL]{downlink}
\acrodef{DOA}[DOA]{direction of arrival}
\acrodef{DoF}[DoF]{degree of freedom}
\acrodef{DPC}[DPC]{dirty-paper coding}
\acrodef{ELAA}[ELAA]{Extra-large aperture array}
\acrodef{ESD}[ESD]{energy spectral density}
\acrodef{FDD}[FDD]{frequency-division duplex}
\acrodef{FSK}[FSK]{frequency-shift keying}
\acrodef{FT}[FT]{Fourier transform}
\acrodef{HT}[HT]{Hilbert transform}
\acrodef{HI}[HI]{harmonic interference}
\acrodef{ICI}[ICI]{inter-carrier interference}
\acrodef{IL}[IL]{insertion losses}
\acrodef{ISI}[ISI]{inter-symbol interference}
\acrodef{IUI}[IUI]{inter-user interference}
\acrodef{JSDM}[JSDM]{joint spatial division and multiplexing}
\acrodef{LBFN}[LBFN]{linear beamforming network}
\acrodef{LINR}[LINR]{leakage-to-interference-plus-noise ratio}
\acrodef{LLF}[LLF]{log-likelihood function}
\acrodef{LMD}[LMD]{linearly modulated digital}
\acrodef{LoS}[LoS]{line-of-Sight}
\acrodef{LSP}[LSP]{Leakage Subspace Precoding}
\acrodef{MAP}[MAP]{maximum a posteriori}
\acrodef{MC}[MC]{Monte Carlo}
\acrodef{MIMO}[MIMO]{multiple-input multiple-output}
\acrodef{MIMOME}[MIMOME]{multi-input, multi-output, multi-eavesdropper}
\acrodef{ML}[ML]{maximum likelihood}
\acrodef{MMSE}[MMSE]{minimum mean squared error}
\acrodef{MMV}[MMV]{multiple measurement vector}
\acrodef{mmWave}[mmWave]{millimeter wave}
\acrodef{MRC}[MRC]{maximum ratio combining}
\acrodef{MRT}[MRT]{maximum ratio trasmission}
\acrodef{MSE}[MSE]{mean squared error}
\acrodef{MUSIC}[MUSIC]{multiple signal classification}
\acrodef{NLOS}[NLOS]{non line-of-sight}
\acrodef{NMSE}{normalized mean squared error}
\acrodef{OFDM}[OFDM]{orthogonal frequency-division multiplexing}
\acrodef{OFDMA}[OFDMA]{orthogonal frequency-division multiple access}
\acrodef{OMP}[OMP]{orthogonal matching pursuit}
\acrodef{OSMP}[OSMP]{orthogonal subspace matching pursuit}
\acrodef{PA}[PA]{power amplifier}
\acrodef{PC}[PC]{partial collusion}
\acrodef{PLS}[PLS]{physical layer security}
\acrodef{PS}[PS]{phase shifter}
\acrodef{PSK}[PSK]{phase-shift keying}
\acrodef{PW}[PW]{planar-wavefront}
\acrodef{QAM}[QAM]{quadrature amplitude modulation}
\acrodef{RF}[RF]{radio frequency}
\acrodef{RFC}[RFC]{Rayleigh fading channel}
\acrodef{SDMA}[SDMA]{space-division multiple access}
\acrodef{SER}[SER]{symbol error rate}
\acrodef{SIC}[SIC]{successive interference cancellation}
\acrodef{SR}[SR]{sideband radiation}
\acrodef{SINR}[SINR]{signal-to-interference-plus-noise ratio}
\acrodef{SLL}[SLL]{side-lobe level}
\acrodef{SOCP}[SOCP]{second-order cone program}
\acrodef{SOMP}[SOMP]{simultaneous-orthogonal matching pursuit}
\acrodef{SPDT}[SPDT]{single-pole-double-throw}
\acrodef{SPST}[SPST]{single-pole-single-throw}
\acrodef{SR}[SR]{sideband radiation}
\acrodef{SS}[SS]{spatial smoothing}
\acrodef{SNR}[SNR]{signal-to-noise ratio}
\acrodef{SUS}[SUS]{successive user selection}
\acrodef{SW}[SW]{spherical-wavefront}
\acrodef{AW}[AW]{successive user selection}
\acrodef{TC}[TC]{total collusion}
\acrodef{TDD}[TDD]{time-division duplex}
\acrodef{TM}[TM]{time modulation}
\acrodef{TMA}[TMA]{time-modulated array}
\acrodef{UL}[UL]{uplink}
\acrodef{ULA}[ULA]{uniform linear array}
\acrodef{UPA}[UPA]{uniform planar array}
\acrodef{UPW}[UPW]{uniform plane wave}
\acrodef{VGA}[VGA]{variable gain amplifier}
\acrodef{VPS}[VPS]{variable phase shifter}
\acrodef{VR}[VR]{visibility regions}
\acrodef{XL}[XL]{extra-large}
\acrodef{ZF}[ZF]{zero-forcing}

\makeatletter
\def\blfootnote{\xdef\@thefnmark{}\@footnotetext}
\makeatother

\begin{document}
	
	\title{\LARGE{Leakage Subspace Precoding \GJAL{and Scheduling} for \\Physical Layer Security in Multi-User XL-MIMO Systems}}
	

\author{Gonzalo~J.~Anaya-L\'opez, Jos\'e~P.~Gonz\'alez-Coma and F.~Javier~L\'opez-Mart\'inez}
	\maketitle
	\blfootnote{\noindent This work has been submitted to the IEEE for possible publication. Copyright may be transferred without notice, after which this version may no longer be accessible. The work of G.~J.~Anaya-L\'opez and F.~J.~L\'opez-Mart\'inez was funded by MCIN/AEI/10.13039/501100011033 through grant PID2020-118139RB-I00, by European Social and Regional Funds and by Junta de Andaluc\'ia (P18-RT-3175, SNGJ5Y6-57, EMERGIA20 00297). The work of J.~P.~Gonz\'alez-Coma was funded by Xunta de Galicia (ED431G2019/01). The authors thank the Defense University Center at the Spanish Naval Academy (CUD-ENM) for the support provided for this research.}
\blfootnote{\noindent G.J.~Anaya-L\'opez and F.J.~L\'opez-Mart\'inez are with the Communications and Signal Processing Lab, Telecommunication Research Institute (TELMA), Universidad de M\'alaga, M\'alaga 29010, Spain. J.P.~Gonz\'alez-Coma is with Defense University Center at the Spanish Naval Academy, Universidad de Vigo, Mar\'i­n 36920, Spain. F.J.~L\'opez-Mart\'inez is also with Dept. Signal Theory, Networking and Communications, Universidad de Granada, Granada 18071, Spain. (contact e-mail: $\rm  gjal@ic.uma.es$.)}
\begin{abstract}
		\textcolor{black}{
We investigate the achievable secrecy sum-rate in a multi-user XL-MIMO system, on which user distances to the base station become comparable to the antenna array dimensions. We show that the consideration of spherical-wavefront propagation inherent to these set-ups is beneficial for physical-layer security, as it provides immunity against eavesdroppers located in similar angular directions that would otherwise prevent secure communication under classical planar-wavefront propagation. A leakage subspace precoding strategy is also proposed for joint secure precoding and user scheduling, which allows to improve the secrecy sum-rate compared to conventional zero-forcing based strategies, under different eavesdropper collusion strategies.}
		%
	\end{abstract}
	
	\IEEEpeerreviewmaketitle
	
	
	\section{Introduction}\label{sec:intro}




\acp{ELAA} or \ac{XL} \ac{MIMO} systems are the natural evolution of massive \ac{MIMO} systems \cite{Carvalho2020}, bringing the opportunity to deploy \acp{BS} with an even larger number of antenna elements to serve a larger set of users. Since antenna sizes in \ac{XL}-\ac{MIMO} deployments become comparable to user distances, new channel features that differ from classical \ac{PW} \cite{LuZe20,LuZe21} and stationarity \cite{Ali2019} assumptions have to be incorporated into the precoding design. One remarkable feature that arises when assuming \ac{SW} propagation is the possibility of separating users in the spatial domain using both the angular direction \textit{and} the distance with respect to the \ac{BS} \cite{LuZe22}.

In the roadmap to defining 6G technology, several new use cases that enhance data security are being proposed \cite{Dang2020}, and \ac{PLS} is identified as a top contender to enable these secure features \cite{Nguyen2021}. The literature of \ac{PLS} is rich when assuming conventional multi-user \ac{MIMO} settings \cite{Geraci2012,Geraci2014}. Different strategies have been proposed for secure precoding design \cite{Maeng2022} and user scheduling \cite{Lee2018} in massive MIMO set-ups, noting that the use of artificial noise may not always be helpful to improve the secrecy rates \cite{Loyka2021}. However, the impact of the inherent channel characteristics of \ac{XL}-\ac{MIMO} systems on \ac{PLS} have not been addressed before, to the best of our knowledge. Hence, in this work we seek to find answers to two key questions: (\emph{i}) \textit{how does \ac{SW} propagation impact the achievable secrecy performance in \ac{XL}-\ac{MIMO} systems?} (\emph{ii}) \textit{how can user scheduling be improved when incorporating security constraints into the \ac{XL}-\ac{MIMO} set-up?} \textcolor{black}{We show that the user decorrelation with distance experienced under \ac{SW} propagation \cite{LuZe22} can be leveraged to reduce information leakage to eavesdroppers}. Thus, if the \ac{BS} successfully exploits this \textcolor{black}{feature}, the achievable secrecy sum-rate substantially increases compared to the conventional \ac{PW} assumption. To that end, we \textcolor{black}{also} propose a novel joint scheduling and precoding scheme that improves the system secure performance compared to state-of-the-art solutions.

\textit{Notation}: Throughout the text, sets are represented with calligraphic font, e.g., $\mathcal{C}$, whereas $|\mathcal{C}|$ denotes the cardinality of a set; uppercase and lowercase bold letters denote matrices and vectors, respectively; $\mathcal{N}_\mathbb{C}(\cdot ,\cdot )$ is the circularly symmetric complex Gaussian distribution; $\mathbb{C}$ is the set of complex numbers;  ${\rm diag}\left ( {\bf{x}} \right )$ is a diagonal matrix with the diagonal given by ${\bf{x}}$;  ${\rm span}\left ( {\mathcal{S}} \right )$ is the span of a set of vectors; $\cap$ and $\subseteq$ are the intersection and subset operators, respectively; $\left ( \cdot \right )^{\rm T}$ denotes transpose; $\left ( \cdot \right )^{\rm H}$ is the Hermitian transpose; $\left[\cdot\right]^+$ denotes $\max\{\cdot,0\}$, where $\max$ represents the maximum value operator; $\argmax\limits_{z} f(z)$ denotes the value of $z$ that maximizes the function $f(\cdot)$.


	\section{System model}\label{sec:model}

Let us consider a general \ac{DL} scenario, where a \ac{BS} equipped with an \ac{XL} antenna array with $M\gg 1$ elements gives service to a set $\mathcal{U}$ of single-antenna users, with $|\mathcal{U}|=K$. We assume an \ac{ULA} as the antenna arrangement for the \ac{BS}, which is placed at the origin of a 2-D plane and deployed along the y-axis. The \ac{BS} operates in two modes \cite{Gonzalo2021}: in \textit{normal} mode, \ac{CSI} for all $K$ users is acquired in an \ac{UL} phase, which is then used to serve the \ac{DL} users according to some sum-rate criterion \cite{gonzalez2021}; in \textit{secure} mode, a set $\mathcal{B}$ of legitimate users (i.e., Bobs) with $|\mathcal{B}|=\kBob$ is served according to some secrecy sum-rate criterion, whereas the remaining set of users $\mathcal{V}$, with $|\mathcal{V}|=\kEve=K-\kBob$, are considered as potential eavesdroppers (i.e., Eves). The data intended to each legitimate receiver potentially experiences an amount of information leakage, which may be used by the eavesdroppers to compromise the transmission \cite{Geraci2012}.

\textcolor{black}{We consider that the array is centered at the coordinate origin and deployed along the vertical axis of a 2-D plane, $r_k$ is the distance from the origin to user $k$, and $\theta_k$ the angle formed by a vector starting at the origin and ending at user's $k$ position.}
The distance between a given user, indexed by $k$, and the $m$-th element of the antenna array is defined as
\begin{align}
	r_{\rm k,m}=r_{\rm k}\sqrt{1-2m{d}_{\rm k}\sin\theta_{\rm k}+{d}_{\rm k}^2m^2},\,m\in\left[-\tfrac{M}{2},\tfrac{M}{2}\right],
	\label{eq:antennaDistance}
\end{align}
\textcolor{black}{so that $r_{\rm k,0}=r_{\rm k}$. Moreover}, $d_{\rm k}=\frac{d}{r_{\rm k}}$ is determined by the separation between two consecutive antenna elements, $d=\frac{\lambda}{2}$, with $\lambda$ being the wavelength. From \eqref{eq:antennaDistance}, we obtain the array response vector for each user as  
\begin{equation}
	{\bf{a}}_{\rm k}=[a_{\rm 1}(r_{\rm k},\theta_{\rm k}),a_{\rm 2}(r_{\rm k},\theta_{\rm k}),\ldots,a_{\rm M}(r_{\rm k},\theta_{\rm k})]^\Tr,
	\label{eq:arrayResponse}
\end{equation}
\GJAL{where the $m$-th element of the vector is expressed, depending on the assumed propagation model, as}
\begin{equation}
	a_m^{\rm SW}(r_{\rm k},\theta_{\rm k})=\tfrac{\sqrt{\beta_0}}{r_{\rm k,m}}\eul^{-\imj\frac{2\pi}{\lambda}r_{\rm k,m}}
	\label{eq:antenaElemSW}
\end{equation}
and
\begin{equation}
	a_m^{\rm PW}(r_{\rm k},\theta_{\rm k})=\tfrac{\sqrt{\beta_0}}{r_{\rm k}} \eul^{-\imj \frac{2\pi}{\lambda} r_{\rm k}} \eul^{-\imj 2\pi m \sin\theta_{\rm k}},
	\label{eq:antenaElemPW}
\end{equation}
respectively, and $\beta_0$ denotes the channel power at the reference distance $r_{\rm ref}=1$ m \cite{LuZe21}.

During the \ac{DL} transmission in \textit{secure} mode, the \ac{BS} sends the data symbols $s_{\rm k} \sim\mathcal{N}_\mathbb{C}(0,1)$ to each legitimate user $k \in \mathcal{B}$, by adapting the transmitted signal ${\bf{x}}$ to the instantaneous channel state acquired in \textit{normal} mode operation through the beamforming matrix ${\bf{W}} \in \mathbb{C}^{M\times \kBob}$, where we define the $k$-th column as ${\bf w}_{\rm k},$ with \mbox{$\|{\bf w}_{\rm k}\|=1, \forall k \in \mathcal{B}$}. Therefore, it follows that the transmitted signal is ${\bf{x_{\rm}}}=\sum_{k \in \mathcal{B}} p_{\rm k} {\bf{w_{\rm k}}} s_{\rm k}$, where $p_{\rm k}$ are the power allocation 
 scale factors such that $P_{\rm TX}=\sum_{k \in \mathcal{B}} p_{\rm k}$, with $P_{\rm TX}$ the \ac{BS} transmit power. 

The received signal ${y_{\rm k}}$ for each legitimate user can be expressed in terms of the previously defined parameters as
\begin{equation}
	y_{\rm k}=\sqrt{p_{\rm k}} s_{\rm k} {\bf{w}}_{\rm k}^\He {\bf{a}}_{\rm k} + \sum_{j \neq k; j \in \mathcal{B}} \sqrt{p_{\rm j}} s_{\rm j} {{\bf w}}_{\rm j}^\He {{\bf a}}_{\rm k} + n_{\rm k},
	\label{eq:receivedSignalu}
\end{equation}
where the first term in \eqref{eq:receivedSignalu} is the desired signal for the legitimate user, the second one is the \ac{IUI} due to legitimate users other than $k$ being served simultaneously, and ${n_{\rm k}}\sim\mathcal{N}_\mathbb{C}(0,\sigma^2_w)$ is the \ac{AWGN}. 
The \ac{SINR} for the legitimate users is defined as \cite{YoGo06}
\begin{equation}
	{\rm{SINR}_{\rm k}}={\frac{p_{\rm k}| {\bf w}_{\rm k}^\He {\bf a}_{\rm k}|^2} {\sigma^2_w + \sum_{ j \neq k; j \in \mathcal{B}} p_{\rm j}| {\bf w}_{\rm j}^\He {\bf a}_{\rm k}|^2}}.
	\label{eq:SINR}
\end{equation}
Now, the legitimate symbols also reach the remaining users $v \in \mathcal{V}$. From the point of view of these users (acting as eavesdroppers), such received signals are information leakage for the message $s_k$, not interference. Therefore, a \ac{LINR} can be defined as ${{\rm LINR}_{\rm k}(v)}={\frac{p_{\rm k}| {\bf w}_{\rm k}^\He {\bf a}_v|^2} {\sigma^2_w}}$, 
which can be seen as the \ac{SINR} associated to the message $s_k$ at the $v$-th eavesdropper. As in \cite{Geraci2012}, we assume the worst-case scenario on which the eavesdroppers have the ability to separate the $k$ symbols (either by means of collusion, or in a genie-aided fashion).

We consider two practical scenarios for eavesdropping: (\textit{i}) \ac{TC}, on which all eavesdroppers $v \in \mathcal{V}$ collaborate into jointly decoding the message targeted to user $k$; (\textit{ii}) \ac{PC}, on which only the eavesdroppers in the vicinity of user $k$ are able to collude, i.e., $v \in \mathcal{E}_k$, where $\mathcal{E}_k$ is the set of colluding clustered eavesdroppers for user $k$. Hence, the overall \ac{LINR} is expressed as \cite{Geraci2014}
\begin{align}
	{{\rm LINR}_{\rm k}^{\rm TC/PC}}=\sum_{v \in \mathcal{V}/\mathcal{E}_k} {\frac{p_{\rm k}| {\bf w}_{\rm k}^\He {\bf a}_v|^2} {\sigma^2_w}}.
	\label{eq:LINR}
\end{align}
The rate of secure information is measured by the secrecy sum-rate, which is given from \eqref{eq:SINR} and \eqref{eq:LINR} as follows \cite{Khisti2010,Geraci2012}:
\begin{equation}
	R_{\rm s} = \sum_{k \in \mathcal{B}} R_{\rm s,k} = \sum_{k \in \mathcal{B}} \log_2 \left[ \frac{1+{\rm SINR}_{\rm k}}{1+{\rm LINR}_{\rm k}^{\rm TC/PC}} \right]^+.
	\label{eq:sumRate}
\end{equation}



	\section{User scheduling and precoding design}\label{sec:pls}

The aim is to maximize the secrecy sum-rate in \eqref{eq:sumRate} subject to the power constraint, i.e.,
\begin{equation}
	\argmax_{\{{\bf{w}}_{\rm k}, p_k\}_{k \in \mathcal{B}}} R_{\rm s}
	\quad\text{s.t.}\quad 
	\sum_{k \in \mathcal{B}} p_k\leq P_\text{TX}.
	\label{eq:problemForm}\vspace{-.75mm}
\end{equation}

We note from \eqref{eq:SINR} and \eqref{eq:LINR} that the achievable secrecy sum-rate is limited by the interference caused by legitimate users, either because it decreases the \ac{SINR} of legitimate users, or because it increases the \ac{LINR} of eavesdroppers. In the \ac{XL}-\ac{MIMO} regime, it is known that \ac{IUI} (and hence information leakage) can be palliated not only by separating users in the angular domain (i.e., through $\theta_k$), but also in the distance domain even for users in the same direction \cite{LuZe21}. Thus, this new feature for interference \textit{and} leakage reduction must be considered in order to jointly design the precoder and scheduling for \ac{PLS} purposes.

There are key differences between conventional joint scheduling and precoding designs \cite{YoGo06,gonzalez2021}, and their secure counterparts targeting a secrecy sum-rate maximization as the ones here addressed: specifically, (\textit{i}) the classical sum-rate metric is simpler, (\textit{ii}) only users selected by the scheduling procedure are considered to compute \ac{IUI}, and (\textit{iii}) information leakage to unscheduled users (i.e., Eves) is ignored. In the sequel, we describe a \ac{LSP} approach for secure joint scheduling and precoding in this scenario, considering two collusion strategies for the eavesdroppers. An schematic view of the proposed scheduling and precoding method is given in Alg. \ref{alg:LSP}. Moreover, the particularities to address both collusion strategies are detailed in the following.

\textit{1) TC scenario}. This case corresponds to the conventional worst-case collusion where all the eavesdroppers share information to decode data of the legitimate users.

According to the \ac{LINR} definition in \eqref{eq:LINR}, to avoid information leakage we propose to define a leakage subspace $\mathfrak{L}$, spanned by the channel vectors of the eavesdroppers, i.e., $\mathfrak{L}=\spann(\B{a}_{v_1},\B{a}_{v_2},\ldots,\B{a}_{v_{\kEve}})$. Now, if the precoders associated to each legitimate user belong to the orthogonal subspace $\mathfrak{L}^\perp$, we ensure that ${{\rm LINR}_{\rm k}^{\rm TC}}=0$ in \eqref{eq:sumRate}. Under this condition, it is simpler to decide which legitimate users will be allocated power. 

\begin{algorithm}[t]
	\caption{Leakage Subspace Precoding}\label{alg:LSP}
	\begin{algorithmic}[1]
		\small
		\STATE  $i \gets 0$, $\mathcal{S}^{(0)}\gets\emptyset$
		\STATE 	\textbf{TC scenario}\\ $\boldsymbol{\Pi}\gets$ Compute the orthogonal projector into $\mathfrak{L}^\perp$\\
				\textbf{PC scenario}\\ $\boldsymbol{\Pi}_{\rm k}\gets$ Compute the orthogonal projector into $\mathfrak{L}_{\rm k}^\perp$, $\forall k\in\mathcal{B}$
		\STATE $q_{\rm k}^{(0)}\gets$ set initial priorities 
		$\forall k\in\mathcal{B}$
		\REPEAT
		\STATE $k^{(i)}\gets\argmax_{k\in\mathcal{B}}(q_{\rm k}^{(i)})$
		\STATE $\mathcal{S}^{(i+1)}\gets\mathcal{S}^{(i)}\cup\{k^{(i)}\}$
		\STATE 	\textbf{TC scenario}\\ $\B{A}^{(i)}\gets[\B{a}_{k^{(1)}}, \B{a}_{k^{(2)}},\ldots, \B{a}_{k^{(i)}}]^T$, with $k^{(j)}\in\mathcal{S}^{(i+1)}$\\
		$\boldsymbol{\omega}_{k^{(j)}}\gets$ Compute TC-ZF precoders \eqref{eq:zf_precodersTC}, $\forall k^{(j)}\in\mathcal{S}^{(i+1)}$\\
		\textbf{PC scenario}\\ $\B{B}^{(i)}_{k^{(j)}}\gets$ Compute using \eqref{eq:BmatricesPC}, $\forall k^{(j)}\in\mathcal{S}^{(i+1)}$\\
		 $\boldsymbol{\omega}_{k^{(j)}}\gets$ Compute PC-ZF precoders \eqref{eq:zf_precodersPC}, $\forall k^{(j)}\in\mathcal{S}^{(i+1)}$
		\STATE $\B{w}_{k^{(j)}}\gets\boldsymbol{\omega}_{k^{(j)}}/\|\boldsymbol{\omega}_{k^{(j)}}\|$, $\forall k^{(j)}\in\mathcal{S}^{(i+1)}$ 
		\STATE $p_k\gets$ Power allocation using waterfilling, $\forall k^{(j)}\in\mathcal{S}^{(i+1)}$
		\STATE $\mathcal{B}\gets\mathcal{B}\setminus\{k\}$
		\STATE $q_{\rm k}^{(i)}\gets$ update user priorities with \eqref{eq:prioritiesUpdates}
		\STATE $i \gets i+1$
		\UNTIL{$\mathcal{B}=\emptyset$ or stopping criterion is met}
		\end{algorithmic}
\end{algorithm}


The iterative procedure starts by setting an initial priority for each of the legitimate users $q_{\rm k}^{(0)}$, $\forall k\in\mathcal{B}$, computed as $q_{\rm k}^{(0)}=\|\boldsymbol{\Pi}\B{a}_{\rm k}\|$,
where $\boldsymbol{\Pi}$ is the orthogonal projector into the subspace $\mathfrak{L}^\perp$. Next, similar to \cite{YoGo06}, we start an iterative procedure such that at the $i$-th iteration the user with higher priority, i.e.,
\begin{equation}
	k^{(i)}=\argmax_{k\in\mathcal{B}}(q_{\rm k}^{(i)})
	\label{eq:userSelection}
\end{equation} 
is chosen as a candidate to receive a certain amount of the available power budget $P_{\rm TX}$. To verify whether including the candidate user $k^{(i)}$ increases the performance metric \eqref{eq:sumRate}, we compute \ac{TC}-\ac{ZF} precoders for the set of users, $\mathcal{S}^{(i)}\subseteq\mathcal{B}$, comprising the previously selected users at iteration $i$ plus the candidate user $k^{(i)}$ as
\begin{equation}
	[\boldsymbol{\omega}_{k^{(1)}}, 
	\ldots, \boldsymbol{\omega}_{k^{(i)}}]=\boldsymbol{\Pi}^*(\B{A}^{(i)})^\He(\B{A}^{(i)}\boldsymbol{\Pi}^*(\B{A}^{(i)})^\He)^{-1},
	\label{eq:zf_precodersTC}
\end{equation}
where $\B{A}^{(i)}=[\B{a}_{k^{(1)}}, \B{a}_{k^{(2)}},\ldots, \B{a}_{k^{(i)}}]^T$, with $k \in \mathcal{S}^{(i)}$. Note that the projector $\boldsymbol{\Pi}$ is included in the computation of the \ac{TC}-\ac{ZF} precoders. Now, for  $k\in\mathcal{S}^{(i)}$ we denote by $\mathfrak{I}_{\bar{k}}^{(i)}=\spann(\B{a}_{j_1}, \B{a}_{j_2},\ldots, \B{a}_{j_{i-1}})$ the subspace spanned by the channels of the $i-1$ legitimate users in $\mathcal{S}^{(i)}$, with $j_1,\ldots,j_{i-1}\neq k$. Hence, the feasible subspace for user $k$ precoding is given by $(\mathfrak{I}_{\bar{k}}^{(i)})^\perp\cap \mathfrak{L}^\perp$. As the spatial characteristics of the channels strongly depend on the propagation model \cite{LuZe21}, 
the \GJAL{interference \textit{and} leakage reduction} provided by the distance in the \ac{SW} model \cite{LuZe21} offers additional flexibility in the precoder design \textcolor{black}{compared to the PW case}. Finally, the precoders for each of the selected legitimate users is $\B{w}_{\rm k}=\boldsymbol{\omega}_{k}/\|\boldsymbol{\omega}_{k}\|$, $\forall k\in\mathcal{S}^{(i)}$ and the power allocation $\B{P}$ is chosen according to conventional waterfilling procedure \cite{YoGo06}, with $P_{\rm TX}=\sum_{k\in\mathcal{S}^{(i)}} p_{\rm k}$. \GJAL{Note that zero power allocation to artificial noise transmission is optimal for \eqref{eq:sumRate} in scenarios without fading or with multiple cooperative eavesdroppers \cite{Loyka2021}, as the scenario under consideration.}

\textcolor{black}{In addition}, at each iteration $i$, it is important to update the priorities 
according to the interference caused by the user selected in previous iterations, that is
\begin{equation}
		q_{\rm k}^{(i)}=\left\|\left(\boldsymbol{\Pi}-\sum\nolimits_{j=1}^{i-1}\B{w}_{\rm k_j}^{(j)}(\B{w}_{\rm k_j}^{(j)})^\He\right)\B{a}_{\rm k}\right\|,
		\label{eq:prioritiesUpdates}
\end{equation}
where $\B{w}_{\rm k_j}^{(j)}$ is the precoder obtained at the iteration $j$ for the selected user $k_j$. The iterative procedure continues incorporating these candidate users as selected users if the new candidate improves the secrecy sum-rate; otherwise, it is discarded and the iterative procedure ends.


\textit{2) Partial colluding scenario}. Here, we consider a more practical situation where only the subset of eavesdroppers $\mathcal{E}_k\subseteq\mathcal{V}$ located on the vicinity of the legitimate user $k\in\mathcal{B}$ are likely to intercept its data, i.e., $\theta_{\rm v}$ and $r_{\rm v}$ ($\forall v\in\mathcal{E}_k$) are similar to $\theta_k$ and $r_k$. These eavesdroppers belonging to $\mathcal{E}_k$ collude to decode the information of legitimate user $k$. Since the set of eavesdroppers able to collude is smaller than in the \ac{TC} case, {this can be taken advantage of} in the precoding design.

As in the previous scenario, we provide a glimpse of the approached procedure in Alg. \ref{alg:LSP}, since it follows the same lines as in the \ac{TC} case. Nevertheless, the legitimate user priorities $q_{\rm k}$ and the precoder design have to incorporate the particularities of \eqref{eq:LINR}. Contrary to \ac{TC}, in the \ac{PC} case each user has its own leakage subspace defined by $\mathfrak{L}_{\rm k}=\spann(\B{a}_{v_1},\B{a}_{v_2},\ldots,\B{a}_{v_{|\mathcal{E}_k|}})$ and $v_1,v_2,\ldots,v_{|\mathcal{E}_k|}\in\mathcal{E}_k$. As a consequence, we only need to guarantee that the precoder for the user $k$ lies in the subspace $\mathfrak{L}_{\rm k}^\perp$ to force the condition ${{\rm LINR}_{\rm k}^{\rm PC}}=0$, thereby simplifying the objective function in \eqref{eq:problemForm}. To that end, we define the orthogonal projector into the subspace $\mathfrak{L}^\perp_{\rm k}$ as $\boldsymbol{\Pi}_{\rm k}$ and set the initial priorities as $q_{\rm k}^{(0)}=\|\boldsymbol{\Pi}_{\rm k}\B{a}_{\rm k}\|$,
such that only the eavesdroppers in $\mathcal{E}_k$ affect whether user $k$ is scheduled or not.
 Then, the iterative procedure starts and the candidate user selection at the $i$-th iteration is performed using \eqref{eq:userSelection}. Notice that the \ac{TC}-\ac{ZF} precoding design in \eqref{eq:zf_precodersTC} is not valid in this scenario, as the leakage subspace is not common to all the users. As such, for each $k\in\mathcal{S}^{(i)}$ we define the matrix
\begin{equation}
	\B{B}^{(i)}_{\rm k}=[\B{a}_{k},\B{a}_{j_1}, \B{a}_{j_2},\ldots, \B{a}_{j_{i-1}},\B{a}_{v_1},\B{a}_{v_2},\ldots,\B{a}_{v_{|\mathcal{E}_k|}}]^T,
	\label{eq:BmatricesPC}
\end{equation}  
where we included the channel vectors for user $k$, for the $i-1$ users in $\mathcal{S}^{(i)}$ such that $j_1,\ldots,j_{i-1}\neq k$, and the ones for the eavesdroppers $v_1,v_2,\ldots,v_{|\mathcal{E}_k|}\in\mathcal{E}_k$. Then, the PC-\ac{ZF} precoder is computed as
\begin{equation}
	\boldsymbol{\omega}_{k}=\left[(\B{B}_{\rm k}^{(i)})^\He\left(\B{B}_{\rm k}^{(i)}(\B{B}_{\rm k}^{(i)})^\He\right)^{-1}\right]_{:,1},
	\label{eq:zf_precodersPC}
\end{equation}
i.e., the first column of the former matrix computation. As such, we ensure that the precoder for each user $k\in\mathcal{S}^{(i)}$ belongs to the intersection of the subspaces $(\mathfrak{I}_{\bar{k}}^{(i)})^\perp\cap \mathfrak{L}_{\rm k}^\perp$. Again, the precoders are normalized by $\B{w}_{\rm k}=\boldsymbol{\omega}_{k}/\|\boldsymbol{\omega}_{k}\|$, $\forall k\in\mathcal{S}^{(i)}$ and 
waterfilling is used for power allocation $\B{P}$. Regarding the update of the priorities in \eqref{eq:prioritiesUpdates}, the same expression can be used but now including the projector $\boldsymbol{\Pi}_{\rm k}$. 

The key difference of this second scenario is that the legitimate user $k$ is only forced to deal with the \ac{LINR} corresponding to those eavesdroppers clustered around him, which in turn are known to be the more detrimental ones for physical layer security \cite{vazquez2018physical}. Therefore, the available \textcolor{black}{degrees of freedom} for precoding \textcolor{black}{design} are larger, in general, than in the \ac{TC} scenario. This additional flexibility will be more effectively exploited under the \ac{SW} model, \textcolor{black}{taking} 
advantage of the \textcolor{black}{interference and leakage reduction} provided by distance.

	\section{Numerical Results}\label{sec:results}
	\begin{table}[t]
	\centering
	\caption{{Simulation parameter settings.}}\label{tab:Sim1}
	\setlength{\tabcolsep}{5pt}
	\def\arraystretch{1.5}
	\begin{tabular}{|l|l|}
		\hline		
		{\textbf{Parameter}} & {\textbf{Value}}  				\\ \hline\hline
		Channel realizations		&	1000 \\ \hline
		\# legitimate users & 		$K_{\rm B}=[10,20]$ 	\\ \hline
		\# eavesdroppers per Bob & 		$k_{\rm e}=[2^{TC}, 6^{PC}]$ 		\\ \hline
		Wavelength & 			$\lambda=0.1249$ m 				\\ \hline
		Distance between antennas & $d=\frac{\lambda}{2}$ m 	\\ \hline
		Number of antennas & 	$M=100$  						\\ \hline
		Transmit \ac{SNR} & 				 	$[0,5,10,15,20,25]$ dB							\\ \hline
		Rayleigh distance &		$ r_{\rm Rayl} = \frac{2 D^2}{\lambda}$ m \\ \hline
		Critical distance \cite{LuZe20} &		$r_{\rm Crit} \approx 9D$ m \\ \hline
		Distance range &	 	$[3\,r_{\rm Crit}, r_{\rm Rayl}]$ m 			\\ \hline
		Angular range &		 	$[-\frac{\pi}{4}, \frac{\pi}{4}]$ rad \\ \hline
	\end{tabular}
\end{table}


We now evaluate the performance of the joint secure precoding and scheduling \ac{LSP} schemes, and investigate the impact of \ac{SW} propagation in the \ac{XL}-\ac{MIMO} regime compared to the case of incorrectly assuming \ac{PW} propagation. \GJAL{In addition, we compare \ac{LSP} with a conventional scheme in the \ac{PLS} literature: ZF precoding with waterfilling \cite{Geraci2012}}. We consider a \ac{BS} with $M=100$ antennas equispaced $d=\lambda/2$, giving service to a set of randomly located users as follows: $\kBob$ legitimate users are deployed randomly along the coverage region described in Table \ref{tab:Sim1} at positions $\theta_{\rm k}$ and $r_{\rm k}$; then, a number of eavesdroppers $k_{\rm e}$ is deployed in the vicinity of each legitimate user at positions $\theta_{\rm e,k}$ and $r_{\rm e,k}$, for $e=1\ldots k_{\rm e}$, so that $K_{\rm E}=K_{\rm B}\cdot k_{\rm e}$. Specifically, one eavesdropper is always located approximately in the same angular direction as user $k$ (i.e., $\theta_{\rm 1,k}=\theta_{\rm k}+\Delta_{\rm k}$, with $\Delta_{\rm k}\sim\mathcal{U}[\pm 0.1^o]$), and $r_{\rm p}$ meters closer to the \ac{BS}. The remaining $e=2\ldots k_{\rm e}$ eavesdroppers are deployed at a distance $r_{\rm q} = r_{\rm Crit}$ m from Bob (i.e., $r_{\rm q}$ is the radius of a protected zone, on which no eavesdroppers are placed close to Bob) and a certain $\theta_{\rm e,k}$.


\newcommand\subDist{.75}
\begin{figure}
	\centering
	\includegraphics[width=\subDist\columnwidth]{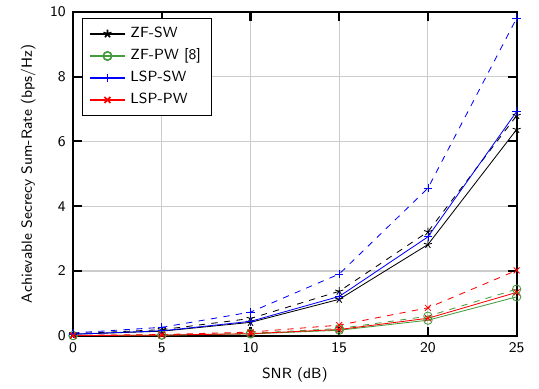}
	\caption{Achievable secrecy sum-rate in the \ac{TC} scenario, considering \ac{SW} and \ac{PW} propagation and different precoding strategies. Solid/dashed lines correspond to ($K_{\rm B}=10$) and ($K_{\rm B}=20$), respectively.}
	\label{fig:ColludingTachievRate}
	\vspace{-2mm}
\end{figure}
\begin{figure}
	\centering
	\includegraphics[width=\subDist\columnwidth]{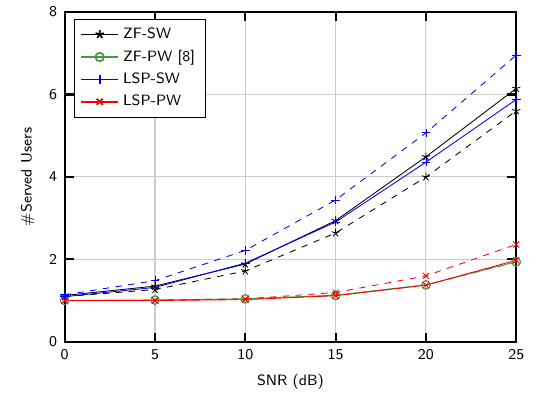}
	\caption{Number of served users in the \ac{TC} scenario, considering \ac{SW} and \ac{PW} propagation and different precoding strategies. Solid/dashed lines correspond to ($K_{\rm B}=10$) and ($K_{\rm B}=20$), respectively.}
	\label{fig:ColludingTservedBobs}
	\vspace{-2mm}
\end{figure}

The achievable secrecy sum-rate (Fig. \ref{fig:ColludingTachievRate}) and the number of served legitimate users (Fig. \ref{fig:ColludingTservedBobs}) are shown for the case of the \ac{TC} scenario, worst-case situation \cite{Geraci2012} where all the eavesdroppers cooperate to decode the legitimate messages, under the \ac{PW} and \ac{SW} propagation models. Solid and dashed lines denote the cases with $K_{\rm B}=\{10,20\}$ legitimate users, respectively, and the remaining set of parameters is listed in Table \ref{tab:Sim1}, with $r_{\rm p}=2r_{\rm q}$ and $D=Md$. 
From the observation of Fig. \ref{fig:ColludingTachievRate}, some remarks are in order: (\emph{i}) under the \ac{PW} model, the secrecy sum-rate is reduced due to the position of the eavesdroppers, and chiefly those located in similar angular directions as the legitimate users; (\emph{ii}) the consideration of the \ac{SW} model allows to improve the secure transmission rates and to serve more users, since legitimate users and eavesdroppers with similar angular directions are decorrelated with distance \cite{LuZe21}; (\emph{iii}) increasing the number of legitimate users notably improves the secrecy sum-rate in the \ac{SW} case, which is not the case when considering \ac{PW}; (\emph{iv}) \textcolor{black}{although the number of served users is not the metric being optimized, we see that in this setting LSP manages to improve \emph{both} the secrecy sum-rate and the average number of served users when $K_{\rm B}=20$, compared to conventional \ac{ZF} precoding \cite{Geraci2012}}.
Note that both \ac{ZF} and \ac{LSP} resort to waterfilling to determine the power allocation. As a consequence, the use of this method is directly related to the number of legitimate users being allocated power. Indeed, users receive power only if their effective channel gain, i.e., the gain obtained when the information leakage and the interference from the other legitimate users are removed, are over a certain threshold, say $\mu$. The value of $\mu$ increases with the number of users receiving power, and decreases with the \ac{SNR} and the effective channel gains. Accordingly, since \ac{ZF} approach gets, in general, more balanced effective gains as users are not prioritized as in \ac{LSP}, we can expect that the number of served users is larger when using  \ac{ZF} compared to \ac{LSP}. This intuition can, however, fail when the \ac{SNR} and/or the effective channel gains are not large enough, as we can see in Fig. \ref{fig:ColludingTservedBobs}. Thus, the increase on $\mu$ due to the number of served legitimate users cannot be compensated due the poor effective channel gains obtained in this scenario.
 
 
\begin{figure}[t]
	\centering
	\includegraphics[width=\subDist\columnwidth]{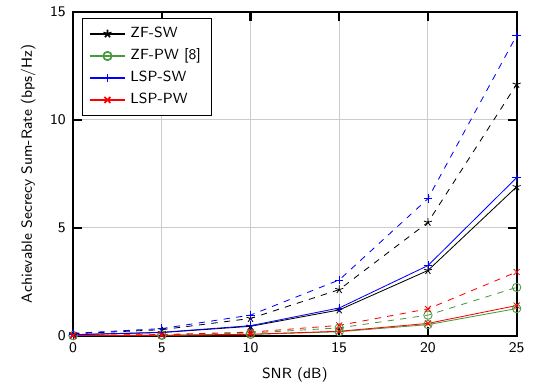}
	\caption{Achievable secrecy sum-rate in the \ac{PC} scenario, considering \ac{SW} and \ac{PW} propagation and different precoding strategies. Solid/dashed lines correspond to ($K_{\rm B}=10$) and ($K_{\rm B}=20$), respectively.}
	\label{fig:ColludingPCachievRate}
	\vspace{-2mm}
\end{figure}
\begin{figure}[t]
	\centering
	\includegraphics[width=\subDist\columnwidth]{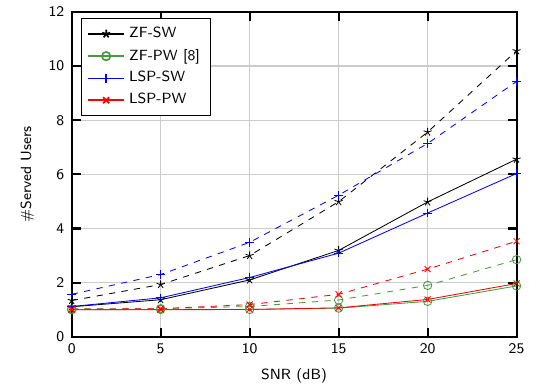}
	\caption{Number of served users in the \ac{PC} scenario, considering \ac{SW} and \ac{PW} propagation and different precoding strategies. Solid/dashed lines correspond to ($K_{\rm B}=10$) and ($K_{\rm B}=20$), respectively.}
	\label{fig:ColludingPCservedBobs}
	\vspace{-2mm}
\end{figure}

The achievable secrecy sum-rate (Fig. \ref{fig:ColludingPCachievRate}) and the number of served legitimate users (Fig. \ref{fig:ColludingPCservedBobs}) are shown for the case of the \ac{PC} scenario, under the \ac{PW} and \ac{SW} propagation models. 
We increase $k_{\rm e}=6$ so that each legitimate user is affected by a larger number of potentially detrimental eavesdroppers compared to the \ac{TC} case. These new eavesdroppers are deployed at a distance $\nicefrac{r_{\rm q}}{2}$ from Bob with the angles $\theta_{\rm e,k}$ uniformly distributed. From the new two figures, Fig. \ref{fig:ColludingPCachievRate} and Fig. \ref{fig:ColludingPCservedBobs}, we extract several remarks: (\emph{i}) under the \ac{PW} model, the secrecy sum-rate is low because close-by eavesdroppers have a dominant effect; (\emph{ii}) this is not the case when considering \ac{SW} propagation, under which the secrecy sum-rate improves by taking advantage of the attenuation of leakage with distance; 
 (\emph{iii}) recall that \ac{ZF} is forced to compute interference cancellation for \emph{all} users, regardless whether they may not be allocated any power in the waterfilling procedure. As the information leakage is not as severe as in the \ac{TC} case, the effective channel gains are larger that in the previous scenario. Therefore, \ac{LSP} prioritizes users with large effective channel gains, whereas \ac{ZF} distributes the power among legitimate users whose effective channel gains are more evenly distributed. Thus, \ac{LSP} achieves better secrecy sum-rates despite serving fewer users, an effect directly related to \ac{SNR}, which is better appreciated in high-SNR regimes.


	\section{Conclusion}\label{sec:conclusion}
	We showed that considering SW propagation in an XL-MIMO system allows to significantly reduce information leakage to eavesdroppers in the same angular directions, compared to \ac{PW} propagation. This new \ac{DoF} allows \ac{SW} to reach higher secrecy-sum rates and to serve more users than in the \ac{PW} case, specially in the worst case scenario on which all eavesdroppers collude, and becomes more beneficial as we increase the number of users. We proposed a novel scheduling and precoding strategy that outperforms conventional zero-forcing strategies in terms of secrecy-sum rate. 


	\bibliographystyle{IEEEtran}
	\bibliography{references}

\end{document}